\documentclass{article}
\newcounter{problem}
\begin{document}

{\center
\section*{Air resistance\footnote{
   It is a pleasure to thank Sanjoy Mahajan for very useful comments.}}
\textsl{David W. Hogg \\
        New York University \\
	\texttt{david.hogg@nyu.edu} \\ 
        2006 September 18 \\ ~}
}

\noindent
Many problems in your book ask you to ignore air resistance; several
even say that you ``can'' ignore air resistance.  Is it true?
Certainly if we say that you \emph{may} ignore air resistance, then
you may.  However, the question remains: When is it \emph{physically
correct} to ignore air resistance?

The detailed analysis of the interaction of air with moving solid
bodies is extremely complex, involving not just pressure but also
wakes, drafts, turbulence, and, at very high speeds, shocks.  However,
it is possible to get reasonable \emph{order-of-magnitude} estimates
of air resistance simply from \emph{dimensional analysis.}

Dimensional analysis is a method for obtaining approximate answers to
physics problems by consideration only (or primarily) of the
\emph{dimensions} or \emph{units} of the quantities involved.  As an
example consider this problem:

\paragraph{Problem~\theproblem}
\refstepcounter{problem} \textsl{A heavy object is dropped from a
height $h$.  How long does it take to hit the ground?  Ignore air
resistance.}

We know (from Galileo's experiments) that all things \emph{not}
strongly affected by air resistance fall with the same acceleration
due to gravity $g$, independent of mass or composition.  So our answer
will depend only on the height $h$ and this acceleration $g$.  The
units of acceleration $g$ are $\mathrm{m\,s^{-2}}$ (meters per second
per second) and the units of the height $h$ is m (meters).  How do we
make a quantity with units of time; in this case s (seconds)?

The acceleration $g$ contains a $\mathrm{s^{-2}}$ but also a pesky m
(meters).  We have to divide out that length with the height $h$.  If
you play around, you will find that the only way to make a time $t$
with an acceleration $g$ and height $h$ is
\begin{equation}
t \sim \sqrt{\frac{h}{g}} \quad,
\label{eq:falltime}
\end{equation}
where I have used a ``$\sim$'' rather than a ``$=$'' symbol because
this answer is not exact in a way I explain below.  Now let's
check the units:
\begin{equation}
\sqrt{\frac{\mathrm{m}}{\mathrm{m\,s^{-2}}}}
= \sqrt{\frac{1}{\mathrm{s^{-2}}}}
= \sqrt{\mathrm{s^2}}
= \mathrm{s}\quad,
\end{equation}
where I have manipulated the units just like one would manipulate
symbols or numbers.

I said that the answer given in (\ref{eq:falltime}) has a ``$\sim$''
symbol and not an ``$=$'' symbol (because it is inexact).  It is
inexact because I have not said what \emph{dimensionless prefactor}
appears in front of the $\sqrt{h/g}$.  For example, the true answer
could be $\pi\sqrt{h/g}$ or $(3/5)\,\sqrt{h/g}$ or any other
dimensionless number times $\sqrt{h/g}$.  Indeed, the true answer has
a factor of $\sqrt{2}$.

Though the answer is wrong in detail, it has many correct features.
For one, it gives you the correct form for the answer (something times
$\sqrt{h/g}$).  For another, it tells you (as you might expect) that
higher heights mean longer free-fall times.  Going further, the
dimensional analysis even tells you that the square of the time is
proportional to the height; that's a pretty detailed result given that
we did almost no thinking to achieve it!

The great thing about physics is that the dimensionless prefactor is
rarely far, far larger or far, far smaller than 1.  Sometimes it is
$\pi$.  Sometimes it is $\pi^2$ (yes, that's close to 10), but rarely
is it 100 or 0.01.

Let's apply this thinking to air resistance.

\paragraph{Problem~\theproblem}
\refstepcounter{problem} \textsl{A soccer ball is kicked hard at
$20\,\mathrm{m\,s^{-1}}$.  Which has a larger magnitude, the
gravitational force on the ball or the air resistance force on the
ball?}

We are only goint to try to get an approximate answer to this problem,
by dimensional analysis.  The magnitude of the gravitational force
$F_\mathrm{g}$ is just $m\,g$ (yes, this is the only combination that
has units of $\mathrm{kg\,m\,s^{-2}}$ (Newtons).  If we recall that
$g\approx 10\,\mathrm{m\,s^{-2}}$ and look up on the web that soccer
balls have masses $m\approx 0.4\,\mathrm{kg}$, we get that
\begin{equation}
F_\mathrm{g} = m\,g \approx 4\,\mathrm{kg\,m\,s^{-2}}\quad.
\end{equation}
How do we estimate the air resistance force?

Air resistance comes, primarily, because there is air in the way of
the ball, and the ball must push the air out of the way.  In detail
this depends on many things, but the dominant dependence is on the
speed $v$ of the ball (because it affects how \emph{quickly} the air
must be moved out of the way), the cross-sectional area $A$ of the
ball (because it affects how \emph{much} air must be moved out of the
way), and the density $\rho$ (``rho'') of the air (because it affects
how much \emph{mass} is associated with that much air).  The speed $v$
has units of $\mathrm{m\,s^{-1}}$ (meters per second), the
cross-sectional area $A$ has units of $\mathrm{m^2}$ (square meters),
and the density $\rho$ has units of $\mathrm{kg\,m^{-3}}$ (kilograms
per cubic meter).  How do we combine these variables to make a
quantity with units of force or $\mathrm{kg\,m\,s^{-2}}$?  Again there
is only one dimensionally correct result:

The magnitude of the air resistance force $F_\mathrm{air}$ is very
well described by
\begin{equation}
F_\mathrm{air}= \xi\,\rho\,A\,v^2\quad,
\end{equation}
where $\xi$ (``xi'') is a dimensionless prefactor (like that discussed
above).  This result is the \emph{only} one possible, given the units
of the quantities at hand!  Checking:
\begin{equation}
\mathrm{kg\,m^{-3}}\,\mathrm{m^2}\,(\mathrm{m\,s^{-1}})^2
= \mathrm{kg\,m^{-3}\,m^2\,m^2\,s^{-2}}
= \mathrm{kg\,m\,s^{-2}}\quad.
\end{equation}

If we guess that $\xi\sim 1$ (not bad for a soccer ball, it turns
out), look up on the web that soccer balls are about $0.25\,\mathrm{m}$
in diameter (or about $0.05\,\mathrm{m^2}$ in cross-sectional area),
and estimate the density of air from what we remember from high-school
chemistry (1 mol of $\mathrm{N_2}$ weighs about $28\,\mathrm{g}$ and
fills about $22\,\mathrm{\ell}$, so it is a little more dense than
$1\,\mathrm{kg\,m^{-3}}$), we find
\begin{equation}
F_\mathrm{air} = \xi\,\rho\,A\,v^2
\sim 20\,\mathrm{kg\,m\,s^{-2}}\quad.
\end{equation}
The force on the soccer ball from the air \emph{is comparable to} the
force of gravity!  Would it be \emph{physically correct} to ignore air
resistance in this case?  No, not at all.

\paragraph{Problem~\theproblem}
\refstepcounter{problem} \textsl{What's the terminal velocity
$v_\mathrm{term}$ for a soccer ball dropped from a tall building?}

If you drop a soccer ball from a tall building, it will accelerate due
to gravity to larger and larger speeds $v$.  Eventually it will
approach the speed at which the air resistance force (which opposes
velocity) balances gravity.  Because we don't know the value of the
dimensionless prefactor $\xi$, we can't answer this question
precisely, but approximately:
\begin{equation}
\rho\,A\,v_\mathrm{term}^2 \sim m\,g
\end{equation}
\begin{equation}
v_\mathrm{term} \sim \sqrt{\frac{m\,g}{\rho\,A}}
\sim 10\,\mathrm{m\,s^{-1}} \quad,
\end{equation}
where we have plugged in numbers from above.

Interestingly, the expression for the terminal velocity can be used to
predict that objects of the same size and shape will have terminal
velocities that vary as the square root of their masses.  This
prediction can be confirmed by comparing the free-fall time of a
basket-shaped coffee filter with that of four stacked basket-shaped
coffee filters (with, therefore, four times the mass of one).

Here are problems to try on your own:

\paragraph{Problem~\theproblem}
\refstepcounter{problem} \textsl{A normal stone brick is dropped from
a height of one meter.  Which has the greater magnitude, the
gravitational force or the air-resistance force?}

Answer this question by assuming that air resistance \emph{doesn't}
matter, and then check the relative magnitudes when the brick is
moving its fastest.  You will find that in this situation ignoring air
resistance \emph{is} physically correct.

That is, there \emph{are} situations in which air resistance
\emph{can} be safely ignored.  The point is that the importance of air
resistance is not a matter of definition or convention: it is a
physical property of the situation that can be checked.

\paragraph{Problem~\theproblem}
\refstepcounter{problem} \textsl{A good soccer player can kick a
soccer ball on an arc that subtends a horizontal distance of
$70\,\mathrm{m}$ (remember the World Cup?).  Use dimensional analysis
to estimate the magnitude of the initial velocity $v$ that the player
must apply to the ball, \emph{ignoring} air resistance.  Now estimate
whether ignoring air resistance was physically correct.  Does a real
player have to kick the ball faster or slower than this speed to make
it go $70\,\mathrm{m}$?}

\paragraph{Problem~\theproblem}
\refstepcounter{problem} \textsl{Same as the previous problem, but for
a golf ball hit $300\,\mathrm{m}$.}

\paragraph{Problem~\theproblem}
\refstepcounter{problem} \textsl{What is the terminal velocity of a US
penny dropped from a tall building?  How does your answer depend on
whether it falls face-on or edge-on?}

\end{document}